\documentclass{llncs}

\usepackage{graphics}
\usepackage{graphicx}
\usepackage{epstopdf}
\usepackage{xspace}
\usepackage{subfigure}
\usepackage{amsmath}
\usepackage{amssymb}
\usepackage{bm}
\usepackage{mathrsfs}
\usepackage{array}
\usepackage{textcomp}
\usepackage{weiwAlgorithm}
\usepackage{url}
\usepackage{balance}
\usepackage{algorithmic}

\newcommand{\eat}[1]{}
\newcommand{\cf}{CF\xspace}
\newcommand{\kclist}{kClist\xspace}
\newcommand{\aot}{AOT\xspace}
\renewcommand{\baselinestretch}{0.955} 

\begin{document}
\title{ AOT: Pushing the Efficiency Boundary of Main-memory Triangle Listing }
\author{ Michael Yu \inst{1} \and Lu Qin \inst{2} \and Ying Zhang \inst{2} \and Wenjie Zhang \inst{1} \and Xuemin Lin \inst{1} }
\institute{University of New South Wales 
\email{\{mryu,wenjie.zhang,lxue\}@cse.unsw.edu.au}
\and University of Technology Sydney 
\email{\{lu.qin,ying.zhang\}@uts.edu.au}}
\maketitle

\begin{abstract}
Triangle listing is an important topic significant in many practical applications.
Efficient algorithms exist for the task of triangle listing.
Recent algorithms leverage an orientation framework, which can be thought of as mapping an undirected graph to a directed acylic graph, namely \textit{oriented graph}, with respect to any global vertex order.
In this paper, we propose an adaptive orientation technique that satisfies the orientation technique
but refines it by traversing carefully based on the out-degree of the vertices in the oriented graph during the computation of triangles. Based on this adaptive orientation technique, we design a new algorithm, namely \aot, to enhance the edge-iterator listing paradigm. We also make improvements to the performance of \aot by exploiting the local order within the adjacent list of the vertices.

We show that \aot is the first work which can achieve best performance in terms of both practical performance and
theoretical time complexity.
Our comprehensive experiments over $16$ real-life large graphs show a superior performance of our \aot algorithm when compared against the state-of-the-art, especially for massive graphs with billions of edges.
Theoretically, we show that our proposed algorithm has a time complexity of
$\Theta(\sum_{ \langle u,v \rangle \in \vec{E} } \min\{ deg^{+}(u),deg^{+}(v)\}))$, where $\vec{E}$ and $deg^{+}(x)$ denote the set of directed edges in an oriented graph and the out-degree of vertex $x$ respectively.
As to our best knowledge, this is the best time complexity among in-memory triangle listing algorithms.

\end{abstract}

\vspace{-4mm}
\section{Introduction}
\label{sec:introduction}
\vspace{-2mm}

Triangle-listing is one of the most fundamental problems in graphs, with numerous applications including structural clustering~\cite{xu2007scan}, web spamming discovery~\cite{becchetti2010efficient}, community search~\cite{berry2011tolerating,radicchi2004defining}, higher-order graph clustering~\cite{yin2017local}, and role discovery~\cite{chou2010discovering}.
A large number of algorithms have been proposed to efficiently enumerate all triangles in a given graph.
These include in-memory algorithms \cite{schank2005finding,07thesistriangle,08tcstriangle,12icdetrikcore}, I/O efficient algorithms \cite{11KDDtriangle,DBLP:journals/tkdd/ChuC12,13sigmodtriangle,14todstrianglelist}, and parallel/distributed algorithms \cite{shun2015multicore,giechaskiel2015pdtl,park2016pte}, etc.
In this paper, we focus on in-memory triangle-listing algorithms and aim to achieve the best performance from both theoretical and practical aspects.

\vspace{1mm}
\noindent \textbf{State-of-the-art.} The existing state-of-the-art in-memory triangle-listing algorithms are based on vertex ordering and orientation techniques \cite{08tcstriangle,danisch2018listing}. Given an undirected simple graph, these algorithms first compute a total-order $\eta$ for all its graph vertices; one example of such ordering is the non-increasing degree order. With the total-order $\eta$, a direction can then be specified for each undirected edge $(u,v)$ such that $\eta(u)<\eta(v)$. Once complete, the graph orientation technique converts the initial undirected graph $G$ into a directed acyclic graph $\vec{G}$. With an oriented graph, the original problem of triangle-listing on an undirected simple graph $G$ is recast to a problem of finding three vertices $u,v,w$ so that directed edges $\langle u,v \rangle$, $\langle v,w \rangle$, $\langle u,w \rangle$ exist in $\vec{G}$. For directed triangle instances, we refer to the role of vertex $u$ with 2 out-going edges that form a triangle as a pivot.

\vspace{1mm}
\noindent \textbf{Motivation.}
We give an example for finding triangles in an oriented-graph \cite{danisch2018listing}.
For each pivot vertex $u$, the method first initializes an index to the out-neighbors of $u$; after that, for each out-neighbor $v$ of $u$, it traverses the out-neighbor $w$ of $v$ and checks to see whether $w$ is also an out-neighbor of $v$ in the index. The advantage of this technique is twofold: First, by simply processing all pivot vertices using the above procedure, it already guarantees that each triangle is generated only once without performing the normal pruning for duplicate triangle solutions. Secondly, parallelization of the algorithm is easy if we process sets of pivot vertices independently. This algorithm has the time complexity of $\Theta(\Sigma_{v\in V} deg^+(v) \cdot deg^-(v))$ and is bounded by $O(m^{1.5})$ \cite{danisch2018listing}, where $V$ is the set of vertices in $\vec{G}$; $deg^+(v)$ and $deg^-(v)$ are the numbers of out-neighbors and in-neighbors of $v$ in $\vec{G}$ respectively; and $m$ is the number of edges in $\vec{G}$.

A dominant cost of the above algorithm is that of look-up operations, where the algorithm searches the out-neighbors of each pivot. A natural question is raised: Is it possible to significantly reduce the number of look-up operations needed by the algorithm? To answer this question,
we first find the time complexity
of the former algorithm is equivalent to that of $\Theta(\Sigma_{\langle u,v\rangle\in \vec{E}} deg^+(v))$, where $\vec{E}$ is the set of directed edges in $\vec{G}$. In other words, for each directed edge $\langle u,v \rangle\in \vec{E}$, the algorithm will always spend $deg^+(v)$ amount of look-up operations irrespective of whether $deg^+(v) \leq deg^+(u)$ holds. We find that if we are able to spend $deg^+(u)$ operations in the case that $deg^+(v) > deg^+(u)$ for the edge $\langle u,v\rangle\in \vec{E}$, the cost of the algorithm can be further lessened, therefore it motivates us to explore new ways to further leverage the properties from graph orientation at a finer level, to improve the algorithm both theoretically and practically.

\vspace{1mm}
\noindent \textbf{Challenges.} Faced with the above problem, we ask intuitively: Can we tackle the asymmetry by manually reversing the direction of each edge $\langle u,v\rangle\in \vec{E}$ if $deg^+(v) > deg^+(u)$ and then reuse the same algorithm on the now modified oriented graph?
Unfortunately, this solution is infeasible. To explain, reversing the direction of the selected edges can result in cycles ($\langle u,v \rangle$, $\langle v,w \rangle$, and $\langle w,u \rangle$) in the oriented graph $\vec{G}$, such cyclic triangles will be missed by the aforementioned algorithm.
To ensure algorithmic correctness, for each undirected edge $(u,v)$, up to two orientations need to be kept simultaneously: the original orientation in $\vec{G}$ and the orientation specified by the comparison of $deg^+(u)$ and $deg^+(v)$ because the two orientations can be inconsistent. Therefore, to make our idea practically applicable, the following issues will be addressed in this paper: (1) How can we integrate the two orientations to improve the algorithm complexity and also ensure that each triangle is enumerated once and only once? and (2) Can we further improve the efficiency of the algorithm practically by exploring some local vertex orders?

\vspace{1mm}
\noindent \textbf{Contributions.} In this paper, we answer the above questions and make the following contributions.

\vspace{1mm}
\noindent  (1) We have designed a new triangle listing algorithm named \textit{Adaptive Oriented Triangle-Listing} (AOT) by developing novel adaptive orientation and local ordering techniques, which can achieve the best
time complexity among the existing algorithms in the literature.

\vspace{1mm}
\noindent (2) We conduct an extensive performance study using 16 real-world large graphs at billion scale, to demonstrate the high efficiency of our proposed solutions. The experimental results show that our AOT algorithm is faster than the state-of-the-art solutions by up to an order of magnitude.
It is also shown that AOT can be easily extended to parallel setting, significantly outperforming the state-of-the-art parallel solutions.

\vspace{1mm}
\noindent \textbf{Organization.}
The rest of the paper is organized as follows. Section~\ref{sec:preli} provides a problem definition and states the notations used and introduce two state-of-the-art methods.
Section~\ref{sec:approach} explains some motivation and explains our proposed algorithm.
Section~\ref{sec:exp} describes the experimental studies conducted and reports on findings from the results.
Section~\ref{sec:rel} presents the related work.
Section~\ref{sec:con} concludes the paper. 
\vspace{-4mm}
\section{Background}
\label{sec:preli}
\vspace{-2mm}

In this section, we formally introduce the problem and the state-of-the-art techniques.
Table~\ref{tb:notations} is a summary of the mathematical notations used in this paper.

\begin{table}[t]
  \centering
  \caption{The Summary of Notations}
  \vspace{-2mm}
  \scalebox{0.9}{
    \begin{tabular}{|c|l|}
      \hline
      \textbf{Notation} & \textbf{Definition} \\ \hline \hline
      $G=(V,E)$ &  an undirected graph with vertices $V$ and edges $E$ \\ \hline
      $\vec{G}$ & a directed graph with vertices $V$ and directed edge $\vec{E}$  \\ \hline
      $u,v,w, x, y$ & vertices in the graph  \\ \hline
      $(u,v)$ & an undirected edge between vertices $u$ and $v$ \\ \hline
      $\langle u,v \rangle$ & a directed edge from vertex $u$ to $v$ \\ \hline
      $(u,v,w)$ & a triangle with vertices $u$, $v$ and $w$\\ \hline
      $deg(u)$ & the degree of the vertex $u$ \\ \hline
      $deg^{+}(u)$ & the out-degree of $u$ in oriented graph \\ \hline
      \end{tabular}
	}
\vspace{2mm}
\label{tb:notations}
\end{table}

\vspace{-2mm}
\subsection{Notations and Problem Definition}
\label{subsec:prob}
\vspace{-1mm}

Let $G=(V, E)$ be an undirected simple graph, where V and E are a set of vertices and a set of edges, respectively.
Below, we also use V(G) and E(G) to denote V and E of a graph G.
The number of vertices and the number of edges is denoted as $n$ and $m$ for $n = |V|$ and $m = |E|$, respectively.
For undirected graph $G$, we denote the set of neighbors of vertex $u$ in $G$ as $N(u)$ and denote the degree of $u$ in $G$ as $deg(u)$ which is equal to $|N(u)|$.
For a directed graph $\vec{G}=(V, \vec{E})$, we use $\vec{E}$ to denote the set of directed edges
$\{ \langle u,v \rangle \}$ where $u$ and $v$ are the starting and ending vertex respectively.
We denote the set of outgoing-neighbors of vertex $u$ in $G$ as $N^{+}(u)$, and the out-degree as $deg^{+}(u) =|N^{+}(u)|$. Likewise, we denote the in-neighborhood of vertex $u$ in $\vec{G}$ as $N^{-}(u)$, and the in-degree as $deg^{-}(u)=|N^{-}(u)|$.
By $(u,v)$, we denote an undirected edge between two vertices $u$ and $v$. A \textbf{triangle} is a set of three vertices fully connected to each other. We denote by $(u,v,w)$ a triangle consisting of three vertices $u$, $v$ and $w$.

\vspace{1mm}
\noindent \textbf{Problem statement.}
Given an undirected simple graph $G=(V,E)$, we aim to develop an efficient main-memory algorithm to
list all triangles in the graph $G$ one by one, with both good time complexity and practical performance.

\vspace{-2mm}
\subsection{Compact Forward (CF) Algorithm}
\label{subsec:cf}
\vspace{-1mm}

We consider the method \textit{Compact-forward} (CF)~\cite{08tcstriangle} as a state-of-the-art for triangle listing; although it was designed in 2008, its efficiency for triangle listing is still referred to frequently~\cite{danisch2018listing}. There are two key components in the \cf algorithm: the ``\textit{edge-iterator}'' computing paradigm and the \textit{orientation} technique.

\vspace{1mm}
\noindent \textbf{Edge-iterator}.
The ``edge-iterator'' is a recurring computing paradigm for triangle listing, its strategy for triangle listing is to find triangles with reference to pairs of adjacent vertices.
Given an edge $(u,v)$, any triangle that includes the edge must contain a third vertex $w$ that has connections to both of $u$ and $v$. Thus, we can obtain any triangles containing edge $(u,v)$ based on the intersection of $N(u)$ and $N(v)$.
For each edge, the edge-iterator returns the set of triangles associated with that edge, and when repeated on all edges, the set of all triangle solutions is made available.

\vspace{1mm}
\noindent \textbf{Orientation technique.}
An orientation technique is recently leveraged in triangle listing algorithms, this involves the generation of a directed (i.e., oriented) graph $\vec{G}$ from an initially undirected input graph $G$~\cite{08tcstriangle}. Each undirected edge is mapped to a directed edge where the direction (i.e., orientation) is decided by the rank of its endpoints in a vertex-ordering
(e.g., out-degree~\cite{08tcstriangle}).
We refer to vertex $u$ as a \textit{pivot vertex} if $u$ has two out-going edges.
We can association a triangle in the undirected graph with only one pivot vertex
to ensure one and only one instance of this triangle in the output,
which significantly improves the performance.

\begin{algorithm}[htb]
\SetVline
\SetFuncSty{textsf}
\SetArgSty{textsf}
\caption{ \textbf{CF($G$)}}
\label{alg:cf}
\Input{
$G$ : an undirected graph
}
\Output{
All triangles in $G$
}
\State{$\vec{G} \leftarrow$ Orientation graph of $G$ based on degree-order}
\label{alg:cf_1}
\For { each vertex $u \in \vec{G}$ }
{
    \label{alg:cf_2}
    \For{ each out-going neighbor $v$ }
    {
        \label{alg:cf_3}
        \State{ $T \leftarrow N^{+}(u) \cap N^{+}(v)$ }
        \label{alg:cf_4}
        \For { each vertex $w \in T$ }
        {
            \State{Output the triangle $(u,v,w)$}
            \label{alg:cf_5}
        }
    }
}
\end{algorithm}

\vspace{1mm}
\noindent \textbf{Compact Forward (CF) Algorithm.} The \cf algorithm is designed based on the edge-iterator and the orientation technique.
We show its pseudocode in Algorithm~\ref{alg:cf}.
In line 1, undirected graph $G$ is transformed into a directed graph $\vec{G}$ via the \textit{orientation} technique.
(Line 2 onward follows the edge-iterator framework.)
In Line 3, triangles are enumerated by iterating through the outgoing-neighborhoods rather than the full neighborhood.
In Line 4, a \textit{merge-based intersection} identifies the common out-going neighbors of $u$ and $v$, denoted by $T$. A set of triangles $(u,v,w)$ is then output for every vertex $w \in T$.

\vspace{1mm}
\noindent \textbf{Analysis.}
Since all triangles identified are unique, a naive traversal of the oriented graphs edges (the outgoing-neighborhoods for each vertex) yields the complete set of unique solutions without explicit duplicate pruning.
In terms of time complexity, the merge-based intersection operation at Line 4 takes $\Theta( deg^{+}(u) + deg^{+}(v))$, assuming that the directed adjacency lists of $u$ and $v$ are sorted.
In total, the \cf algorithm has a complexity of $\Theta( \sum_{ \langle u,v \rangle \in \vec{E} } deg^{+}(u) + deg^{+}(v))$.

\begin{remark}
There is also an alternative implementation of the CF algorithm that adopts hash tables for the intersection operation,
namely CF-Hash. Suppose a hash table has been built for each vertex based on the out-going neighbors in the oriented graph.
At Line $4$ of Algorithm~\ref{alg:cf}, we may choose the vertex with larger number of neighbors
as the hash table for intersection operation with $\Theta(\min\{ deg^{+}(u),deg^{+}(v)\})$ look-up cost.
This can come up with a better time complexity of $\Theta(\sum_{ \langle u,v \rangle \in \vec{E} } \min\{ deg^{+}(u),deg^{+}(v)\}))$.
However, as reported in~\cite{08tcstriangle,shun2015multicore} and our experiments, the practical performance of hash-based CF algorithm is not competitive compared to the above merge-based CF algorithm. Thus, the merge-based CF algorithm is used as the representative of CF algorithm in the literature.
\end{remark}

\vspace{-2mm}
\subsection{k-Clique Listing Algorithm (kClist)}
\label{subsec:kclist}
\vspace{-1mm}

We introduce the second state-of-the-art algorithm for in-memory triangle listing.
The \kclist algorithm~\cite{danisch2018listing} lists cliques for a queried size $k$, we restrict our discussion to the relevant use-case when $k=3$ for listing triangles.
\kclist follows the node-iterator triangle listing paradigm which is described below.

\vspace{1mm}
\noindent \textbf{Node-iterator}.
The ``node-iterator'' triangle listing paradigm lists triangles by inspecting for adjacency between vertex pairs within one vertex neighborhood. For example, consider the neighboring vertices of node $u$, if there is an edge between two neighbors $v_2$ and $v_3$, then the triangle solution (u, $v_2$, $v_3$) is output.

\begin{algorithm}[t]
\SetVline
\SetFuncSty{textsf}
\SetArgSty{textsf}
\caption{ \textbf{kClist($G$)}}
\label{alg:kcl}
\Input{
$G$ : an undirected graph
}
\Output{
All triangles in $G$
}
\State{$\vec{G} \leftarrow$ Orientation graph of $G$ based on degeneracy order $\eta$ }
\label{alg:kcl_1}
\For { each vertex $u \in \vec{G}$ }
{
    \label{alg:kcl_2}
    \For{ any two out-going neighbors $\{v, w\}$ of $u$  with $\eta(v) < \eta(w)$ }
    {
        \label{alg:kcl_3}
        \If{ there is a directed edge $\langle v,w \rangle \in \vec{E}$}
        {
            \label{alg:kcl_4}
                \State{Output triangle $(u,v,w)$}
                \label{alg:kcl_6}
        }
    }
}
\end{algorithm}

\vspace{1mm}
\noindent \textbf{k-Clique Listing (kClist) Algorithm}.
The kClist algorithm begins by generating an oriented graph $\vec{G}$ based on the degeneracy ordering~\cite{batagelj2002generalized}.
We use $\eta$ to denote the degeneracy ordering here.
In lines 3 - 5 of Algorithm~\ref{alg:kcl}, for every two out-going neighbor $v$ and $w$ where $\eta(v) < \eta(w)$,
the existence of a directed edge $\langle v,w \rangle$ is assessed; for each edge found, a triangle solution is output.

\vspace{1mm}
\noindent \textbf{Analysis}.
The running time of kClist is $ \Theta (m + \sum_{u \in V } deg^{+}(u) \times deg^{-}(u)) $, this can also be expressed as $ \Theta( \sum_{ \langle u,v \rangle \in \vec{E} } deg^{+}(v))$. It is apparent that the time complexity of kClist is an improvement compared to the \cf algorithm which takes $\Theta( \sum_{ \langle u,v \rangle \in \vec{E} } (deg^{+}(u) + deg^{+}(v)))$ time, its practical performance is also shown to be efficient.

\section{Our Approach}
\label{sec:approach}

We introduce our adaptive orientation technique and implement it in our algorithm, \aot, to push the efficiency boundary for main-memory triangle listing algorithms.

\vspace{-2mm}
\subsection{Motivation and Problem Analysis}
\label{subsec:motivation}
\vspace{-1mm}

Since the proposal of the orientation technique, its nice properties and good practical performance have allowed it to gradually become a valuable technique utilized in subsequent studies of triangle listing.

\vspace{0.5mm}
\noindent \textbf{Shortcoming of Orientation Technique.}
We recognize the prevalent usage of this orientation technique, however, we respond by showing that although current methods leverage the salient benefits of orientation, there are still finer benefits that are overlooked. We argue that there are still ways to further leverage properties that can improve the existing performances of triangle-listing. Our goal is therefore to maximize the benefits of the orientation technique.

In the following discussion of oriented edges, relative to a vertex $u$, we refer to edges $\langle u, v \rangle$ as \textit{positive edges} if the out-degree relation of its adjacent vertex $v$ has an out-degree value that is greater or equal to that of the pivot vertex ($deg^{+}(u) \leq deg^{+}(v)$); we also refer to edges $\langle u, v \rangle$ as \textit{negative edges} if the pivot vertex has the greater out-degree value ($deg^{+}(u) > deg^{+}(v)$).

\begin{figure}[t]
	\center{\includegraphics[width=0.5\columnwidth]
	{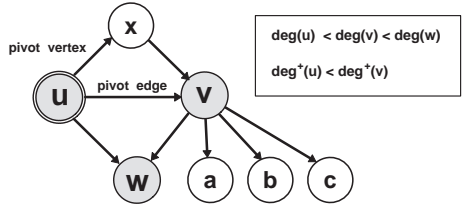}}
	\vspace*{-4mm}
	\caption{Limit of orientation technique}
\vspace{2mm}
\label{fig:limit_orientation}
\vspace{-1mm}
\end{figure}

We refer to the time complexity of the \kclist algorithm and see that it is $\Theta( \sum_{ \langle u,v \rangle \in \vec{E} } deg^{+}(v))$ when listing triangles.
However, this is not optimal for triangle listing.
In Figure~\ref{fig:limit_orientation}, consider the point in the triangle listing computation where triangles of edge $\langle u, v \rangle$ are processed. With respect to pivot vertex $u$, $deg^{+}(v)$ is larger than that of $deg^{+}(u)$ since $4 > 3$. The aforementioned complexity is not favorable for processing this ordinary edge. Its issue is because its cost is strictly that of $deg^{+}(v)$ (i.e., 4). Our observation is that, if we can reverse the direction of the edge $\langle u,v \rangle$ to follow the out-degree vertex-order, we can process $\langle u, v \rangle$ more favorably with $deg^{+}(u)$ (i.e., 3) and come up with a better time complexity.

\vspace{1mm}
\noindent \textbf{Solutions are non-trivial.}
Obviously, the optimal instance for a running time of $\Theta( \sum_{ \langle u,v \rangle \in \vec{E} } deg^{+}(v))$ is an oriented graph $\vec{G}$ where all edges are positive.
However, most graphs do not exhibit that property: in most cases, not all graphs edges $\{ \langle u,v \rangle \}$ are necessarily positive at the same. 
Recall Figure~\ref{fig:limit_orientation}: while edges such as $ \langle u,x \rangle$ and $\langle u,w \rangle$ are positive, negative edges such as $\langle u,v \rangle$ are also possible.

We remedy the existence of negative edges by making a series of modifications to the computed orientation of negative edges after it is oriented. One naive way of achieving this is to manually change the direction of the oriented edge. For example, there is a negative edge $\langle u,v \rangle$ in Figure~\ref{fig:limit_orientation}, we see that it can become a positive edge if its direction is simply changed to $\langle v,u \rangle$. This methodology is limited because it ultimately undermines the total order used in the orientation, moreover, changing $\langle u,v \rangle$ creates a cycle subgraph $(u,x,v,u)$; this is a critical complication since triangle $(u,x,v)$ would surely be omitted and missing from the result set of existing methods.

Ultimately, the out-degree of a vertex is a result of the orientation of its incident edges, and therefore depends directly on the total ordering used for the orientation techniques, it is difficult to significantly reduce the number of negative edges by manually changing its orientation.

\vspace{1mm}
\noindent \textbf{The main idea.}
As an alternative, we instead suggest modifying the computing order of $u$ and $v$ on the fly when encountering negative edges instances $\{ \langle u,v \rangle \}$.
We notice that the \cf algorithm cannot take advantage of this because its complexity of
$\Theta(deg^{+}(u) + deg^{+}(v))$ for every edge $\langle u, v \rangle$ suggests that
the design of \cf is insensitive to the direction of the edge.
We also notice that the \kclist algorithm cannot do this either, because the accessing order of the vertices has to strictly follow the degeneracy order on the oriented graph to ensure the correctness of the algorithm. We have showed that two state-of-the-art techniques cannot trivially take advantage of this observation. In contrast, our algorithm does not depend on any total order, any total order will be acceptable.

Following the above analysis, we are motivated to develop a technique that tightens the boundary for efficient triangle listing, by taking advantage of the resulting out-going degree order of each incoming edge, and adaptively listing triangle based on its property.
Our key idea involves selecting the optimal pivot vertices for each triangle accordingly, such that each triangle is found only from the vertex with a smaller out-going degree.
This way we achieve the time complexity $\Theta( \min\{ deg^{+}(u), deg^{+}(v)) )$ for every edge $\langle u, v \rangle \in \vec{E}$, which is now \textit{optimal} for a given oriented graph following the edge-iterator paradigm.

When finding the intersection between the out-going neighbors of adjacent vertices $u$ and $v$ (i.e. $N^{+}(u) \cap N^{+}(v)$ for an oriented edge $\langle u, v \rangle$), there is a larger and a smaller out-degree vertex, we use the hash-join approach as the appropriate method to perform the set-intersection. Note that one hash-table here contain the out-going neighbors for one single vertex.
Following the hash-join approach, we choose to look-up the out-going neighbors of the vertex with the lesser out-degree in the hash structure of the vertex with the greater out-degree.
However, with one endpoint fixed, within its adjacent neighborhood, the endpoint with the lesser out-degree vertex varies, to accomplish the former statement efficiently is hard because it is not known in advance which endpoint has the smaller out-degree vertex.
The known solution requires both hash tables for either endpoints be available when the edge is visited. There are two methods for constructing the two indexes in advance:
(1) Building hash tables of all graph vertices prior to listing. Where this is a naive solution, it is computationally infeasible due to its high storage demand and a high look-up cost.
(2) For each vertex $u$, building a hash table for all for its out-going neighbors on the fly. This is also infeasible because one vertex is likely to need to rebuild its hash table multiple times throughout the listing stage.

\begin{figure}[t]
	\center{\includegraphics[width=0.7\columnwidth]
	{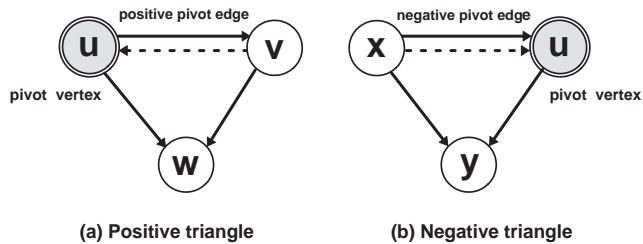}}
	\vspace{-2mm}
	\caption{ Motivation for Adaptive Orientation }
\label{fig:motivation}
\end{figure}

\vspace{1mm}
\noindent \textbf{Categorizing Triangles.}
To facilitate understanding of technique, we discuss two categorizes for each triangle in an oriented graph $\vec{G}$: \textit{positive triangle} or \textit{negative triangle}.
The category of each triangle depends on its pivot edge: given an oriented triangle, were refer to it as positive if its pivot edge $\langle x, y \rangle$ is positive i.e. $\eta(x) < \eta(y)$ and $deg^{+}(x) \geq deg^{+}(y)$; otherwise, it is negative if $\eta(x) < \eta(y)$ and $deg^{+}(x) < deg^{+}(y)$, where $\eta$ is the vertex ordering used in the orientation.

An example instance of positive and negative triangles is shown in Figures~\ref{fig:motivation}(a)-(b). We consider the two triangles from a sample graph with a common vertex $u$, we note that without additional structural information from the graph, the induced subgraphs u,v,w and x,u,y are isomorphic. However, with attention to pivot edges $\langle u, v \rangle$ and $\langle x, u \rangle$, we observe that the two triangles are different in terms of the out-going degree order (the dotted line points to the vertex with the higher out-degree), and there for their orientation is different.

We propose seperate computations for the two types of triangles due to their subtle differences, by selecting different respective piviot veritices. The selection of the pivot vertex affects the amount of computation to list the triangle. If it is a positive triangle, we remain consistent with the orientation technique and use the vertex with two out-going edges as the \textit{pivot vertex}
(e.g. the vertex $u$ in Figure~\ref{fig:motivation}(a)).
However, if it is a negative triangle, we select a different vertex as the pivot vertex (e.g. $u$ in Figure~\ref{fig:motivation}(b)).

A direct benefit from the above selection is that, every vertex $u \in \vec{G}$ is now the pivot vertex for both \textit{positive} and \textit{negative} triangles solutions, where the previous technique rigidly processes all triangles as positive triangles. In the traditional orientation technique, both triangles would be processed equally and listed by vertices $u$ and $x$ respectively, this was because the pivot vertex of each triangle is strictly the vertex with two out-going edges and does not account for our positive or negative triangle definitions.
A desirable property of our \textit{adaptive orientation technique} is that this way each vertex $u$ only needs to build a hash table once for its out-going neighbors.

To conclude:
For positive triangles with pivot vertex $u$,
for each out-going neighbor $v$ of $u$ (e.g. $v$ in Figure~\ref{fig:motivation}(a)),
we will look-up if each out-going neighbors $w$ of $v$ (e.g. $w$ in Figure~\ref{fig:motivation}(a))
is also in the hash table, to see if $w$ is also an out-going neighbor of $u$.
For the negative triangles with pivot vertex $u$,
for each in-coming neighbor $x$ of $u$ (e.g. $x$ in Figure~\ref{fig:motivation}(b)),
we will look-up if each out-going neighbor $y$ of $x$ (e.g. $y$ in Figure~\ref{fig:motivation}(b))
is also in the hash table, to see if $y$ is also an out-going neighbor of $u$.

As we later show in the theoretical analysis in Section~\ref{subsec:approach},
our proposed \textit{adaptive orientation} technique achieves the time complexity of $\Theta( \sum_{ \langle u,v \rangle \in \vec{E} } \min\{ deg^{+}(u),deg^{+}(v)\}))$  because, in terms of the hash-based intersection, the look-up operations will always be performed on the vertex with larger out-degree values for each oriented edge.

\vspace{-2mm}
\subsection{The Algorithm}
\label{subsec:approach}
\vspace{-1mm}

We introduce the algorithm with our proposed adaptive orientation technique.
With reference to the pseudo-code in Algorithm~\ref{alg:hm},
In line 1, the orientated graph $\vec{G}$ is generated following the degree vertex order.
In lines 2-13, a vertex $u$ acts as pivot vertex and lists its associated positive and negative triangles.
For each pivot vertex $u$, Line 3 generates a bitmap hash table $H$ based on its adjacency neighborhood.

For pivot vertex $u$, all \textit{positive triangles} are enumerated in Lines 4-8.
That is, for each out-going neighbor $v$ of $u$ with $deg^{+}(v) < deg^{+}(u)$ (i.e. positive pivot edge),
we find its out-going neighbors which are also out-going neighbors of $u$ by looking up the hash table $H$
as illustrated in Figure~\ref{fig:motivation}(a).

Similarly, all \textit{negative triangles} for pivot vertex $u$ are enumerated in Lines 9-13.
For each in-coming neighbor $x$ of $u$ with $deg^{+}(x) < deg^{+}(u)$ (i.e. negative pivot edge),
we find its out-going neighbors which are also out-going neighbors of $u$ by looking up the hash table $H$ as illustrated in Figure~\ref{fig:motivation}(b).

\begin{algorithm}[t]
\caption{Our Algorithm -- AOT (G)}
\SetVline
\SetFuncSty{textsf}
\SetArgSty{textsf}
\label{alg:hm}
\Input{
	$G$ : an undirected graph
}
\Output{
	All triangles in $G$
}
\State{$\vec{G} \leftarrow$ orientation graph of $G$ based on degree-order}
\label{alg:hm_1}
\For{$u \in V(\vec{G})$}{
    \label{alg:hm_2}
    \State{Set-up the hash table $H$ with IDs of the out-going neighbors of $u$ ($N^{+}(u))$ }
    \label{alg:hm_3}

    \For{every out-going neighbor $v$ of $u$}{
       \label{alg:hm_4}
       \If{$deg^{+}(v) < deg^{+}(u)$}{
         \label{alg:hm_5}
         \For{ every out-going neighbor $w$ of $v$ }{
             \label{alg:hm_6}
             \If {Find $w$ in $H$}{
                \label{alg:hm_7}
                \State{ output triangle ($u,v, w$)}
                \label{alg:hm_8}
             }
         }
      }
   }

    \For{every in-coming neighbor $x$ of $u$ }{
       \label{alg:hm_9}
       \If{$deg^{+}(x) < deg^{+}(u)$}{
         \label{alg:hm_10}
         \For{ every out-going neighbor $y$ of $x$ }{
             \label{alg:hm_11}
             \If {Find $y$ in $H$}{
                \label{alg:hm_12}
                \State{ output triangle ($u,x, y$)}
                \label{alg:hm_13}
             }
         }
      }
   }
}
\end{algorithm}

\vspace{1mm}
\noindent \textbf{Correctness.}
To explain the correctness of our algorithm, we recall that each oriented triangle in $\vec{G}$ belongs to either a positive type triangle or a negative type triangle, we note that this is true for any vertex total-order.

Given an oriented triangle $(u,v,w)$: such that $u$ is the pivot vertex, and $\langle u, v \rangle$ is its pivot edge as illustrated in Figure~\ref{fig:limit_orientation}.
If $deg^{+}(v) < deg^{+}(u)$, then $(u,v,w)$ is a positive triangle with pivot vertex $u$; given $w$ is the common out-going neighbor of $u$ and $v$, a triangle will be output at Line 8 of Algorithm~\ref{alg:hm}.
Otherwise, if the triangle is not positive i.e., if $deg^{+}(u) < deg^{+}(v)$\footnote{Recall that ties are broken by vertex ID.}, $(u,v,w)$ is a negative triangle with pivot vertex $v$, this oriented triangle will be output at Line 13 of Algorithm~\ref{alg:hm} when $v$ is the pivot vertex, because $w$ is the common out-going neighbor of $u$ and $v$.
Evidently, this oriented triangle $(u,v,w)$ will not be output under any other scenario when following the oriented triangle technique. Consequently, this showed that $(u,v,w)$ will be output once and only once, the correctness of the Algorithm~\ref{alg:hm} follows.

\vspace{1mm}
\noindent \textbf{Time Complexity.} 
We use a bitmap with size $|V|$ to implement the hash table $H$,
where $H[v.ID]=1$ if the vertex $v$ is the out-going neighbor of the pivot vertex.
For each pivot vertex $u$ visited, we can use $\Theta(deg^{+}(u))$ time to initiate or
clean the hash table $H$. Thus, the maintenance of $H$ takes $\Theta(2m)$ time.

Recall that for a pivot edge $\langle u,v \rangle$, the set of triangles it outputs can be a mix of either positive or negative triangles. For every pivot edge $\langle u,v \rangle$,
the time complexity for its positive triangles is $\Theta(deg^{+}(v))$ with $deg^{+}(v) < deg^{+}(u)$
since the time complexity of Line 8 is $\Theta(1)$.
Similarly, the time complexity for its negative triangles is $\Theta(deg^{+}(u))$ with $deg^{+}(u) < deg^{+}(v)$ since the time complexity of Line 13 is $\Theta(1)$.
It follows that the total time complexity of our Algorithm~\ref{alg:hm} is $\Theta(\sum_{ \langle u,v \rangle \in \vec{E} } \min\{ deg^{+}(u),deg^{+}(v)\}))$

\begin{figure}[t]
	\center{\includegraphics[width=0.4\columnwidth]
	{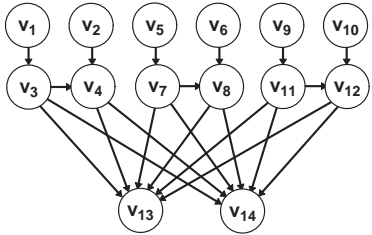}}
	\vspace{-2mm}
	\caption{ Example Graph }
\label{fig:example_time}
\end{figure}

\begin{example}
In Figure ~\ref{fig:example_time}, the oriented graph has 14 vertices and 21 edges. Out of the 21 edges, 9  for which have a $deg^{+}(v)$ value of greater than 0.
\vspace{2mm}
For $\sum_{ \langle u,v \rangle \in \vec{E} } deg^{+}(v))$:
Edges $\langle v_1 , v_3 \rangle$,
$\langle v_5 , v_7 \rangle$ and
$\langle v_9 , v_{11} \rangle$ each incur a cost of 3.
Edges $\langle v_2 , v_4 \rangle$,
$\langle v_6 , v_8 \rangle$,
$\langle v_{10} , v_{12} \rangle$ each incur a cost of 2.
Edges $\langle v_3 , v_4 \rangle$,
$\langle v_7 , v_8 \rangle$ and
$\langle v_{11} , v_{12} \rangle$ each also incur a cost of 2.
The remaining edges incur no cost.
In total, $\sum_{ \langle u,v \rangle \in \vec{E} } deg^{+}(v))$ $= 3 + 3 + 3 + 2 + 2 + 2 + 2 + 2 + 2 = 21$.
\vspace{2mm}
For $\sum_{ \langle u,v \rangle \in \vec{E} } \min\{ deg^{+}(u),deg^{+}(v)\})$:
Edges $\langle v_1 , v_3 \rangle$,
$\langle v_5 , v_7 \rangle$ and
$\langle v_9 , v_11 \rangle$ each incur a cost of 1.
Edges $\langle v_2 , v_4 \rangle$,
$\langle v_6 , v_8 \rangle$,
$\langle v_{10} , v_{12} \rangle$ each also incur a cost of 1.
Edges $\langle v_3 , v_4 \rangle$,
$\langle v_7 , v_8 \rangle$ and
$\langle v_{11} , v_{12} \rangle$ each incur a cost of 2.
The remaining edges incur no cost.
In total, $\sum_{ \langle u,v \rangle \in \vec{E} } \min\{ deg^{+}(u),deg^{+}(v)\})$
$= 1 + 1 + 1 + 1 + 1 + 1 + 2 + 2 + 2 = 12$.

The former is a calculation of the computation required by the state-of-the-art, the latter is the computations required by our algorithm. In comparison, Example 1 illustrates that our algorithm incurs significantly fewer computation to list triangles. Where the costs for some edges is the same between two algorithms, our algorithm uses less computation for edges $\langle v_1 , v_3 \rangle$,
$\langle v_5 , v_7 \rangle$,
$\langle v_9 , v_{11} \rangle$,
$\langle v_2 , v_4 \rangle$,
$\langle v_6 , v_8 \rangle$ and
$\langle v_{10} , v_{12} \rangle$.
\end{example}

\begin{remark}
\label{remark:bitmap}
Note that the bitmap hash table cannot be deployed by CF-Hash technique.
Clearly, on large graphs we cannot afford to construct $|V|$ bitmap hash tables each of which has size $|V|$.
On the other hand, it is time consuming to build $H$ on the fly because,
unlike we build the hash table $H$ only once\footnote{When it is chosen as the pivot vertex} for each vertex in AOT algorithm, $H$ might be built multiple times for a vertex because it's hash table will be chosen (i.e., build on the fly)
by CF-Hash algorithm once its hash table size is smaller than that of pivot vertex.
\end{remark}

\vspace{1mm}
\noindent \textbf{Space Complexity.}

We only need to keep the graph $G$, the oriented graph $\vec{G}$ and the global hash table $H$, as a result, Algorithm~\ref{alg:hm} is space efficient, with space complexity $O(m+n)$ where $m$ is the number of edges
and $n$ is the number of vertices in $G$.

\vspace{1mm}
\noindent \textbf{Exploiting Local Order.}
In addition to the global vertex order, we also consider a local vertex ordering used to store vertices within the scope for each vertex neighborhood list (i.e. \textit{local order}).
In Algorithm~\ref{alg:hm}, the dominant cost is the hash table look-ups happen at Lines 7 and 12.
There is a good chance that a vertex $w$ will be repeatedly checked because of the overlap of the neighborhood.
Ideally, the ID of $w$ should be kept in the CPU cache such that the following look-up of $w$ can be processed efficiently.
We may design sophisticate local ordering strategy with some assumptions on the workload such that
the neighbors of a (pivot) vertex is well organized by their neighborhood similarity.
However, we cannot afford such cost for the preprocessing. In this paper, we order the vertices in the adjacent list of a vertex by the decreasing order of their degree;
that is, we visit the vertices at Lines 4 and 9 in Algorithm~\ref{alg:hm}
following the degree order.
This is because we believe the vertex with a high degree is more likely to have common neighbors with other vertices.
For each vertex, we can keep its neighbors with this local order in the adjacent list. Our empirical study confirms the efficiency of this local order strategy.


\vspace{-4mm}
\section{Experimental Study}
\label{sec:exp}
\vspace{-2mm}

\begin{table*}[t]
\center
\begin{tabular}{|l|l|l|l|l|l|}
\hline
Graph      & \#Nodes (M) & \#Edges (M) & Avg. Degree & Max. Degree  & \#Triangles (M)\\
\hline
web-baidu-baike     & 2.14  & 17.01    & 8  & 97,848    & 25.21     \\
uk-2014-tpd         & 1.77  & 15.28    & 9  & 63,731   & 259.04    \\
actor               & 0.38  & 15.04    & 39 & 3,956   & 346.81    \\
flicker             & 1.62  & 15.48    & 10 & 27,236   & 548.65    \\
uk-2014-host        & 4.77  & 40.21    & 8  & 726,244 & 2,509.74  \\
sx-stackoverflow    & 6.02  & 28.18    & 5  & 44,065   & 114.21    \\
ljournal-2008       & 5.36  & 49.51    & 9  & 19,432   & 411.16    \\
soc-orkut           & 3.00  & 106.35   & 35 & 27,466   & 524.64    \\
hollywood-2011      & 2.18  & 114.49   & 53 & 13,107 & 7,073.95  \\
indochina-2004      & 7.41  & 150.98   & 20 & 256,425 & 60,115.56 \\
soc-sinaweibo       & 58.66 & 261.32   & 4  & 278,489   & 212.98 \\
wikipedia\_link\_en & 12.15 & 288.26   & 24 & 962,969 & 11,686.21 \\
arabic-2005         & 22.74 & 553.90   & 24 & 3,247 & 36,895.36 \\
uk-2005             & 39.46 & 783.03   & 20 & 5,213   & 21,779.37 \\
it-2004             & 41.29 & 1,027.47 & 25 & 9,964 & 48,374.55 \\
twitter-2010        & 41.65 & 1,202.51 & 29 & 2,997,487 & 34,824.92 \\
\hline
\end{tabular}
\vspace{2mm}
\caption{Statistics of 16 Datasets.}
\label{table:data}
\end{table*}

\vspace{1mm}\noindent \textbf{Algorithms.}
To show the efficiency of our proposed technique, we compare our proposed algorithm with the following state-of-the-art methods. In total, we make comparisons between the four algorithms listed below.

\begin{itemize}
\item \textbf{CF} ~\cite{08tcstriangle,shun2015multicore}. The \cf algorithm, presented in Section~\ref{subsec:cf}.
\item \textbf{CF-Hash}~\cite{08tcstriangle,shun2015multicore}. A variant of CF, where the intersection of two adjacency lists are implemented by hashing.
\item \textbf{kClist} ~\cite{danisch2018listing}. The kClist algorithm for triangle listing, presented in Section~\ref{subsec:kclist}.

\item \textbf{AOT}. Our proposed algorithm with adaptive orientation and local ordering, presented in Section~
\ref{subsec:approach}.
\end{itemize}

The source-code for the assessment of $CF$, $kClist$, and $CF$-$Hash$ are acquired from their respective authors. We note that for $CF$ and $CF$-$Hash$, we use the implementation from~\cite{shun2015multicore} named $TC$-$Merge$ and $TC$-$Hash$ respectively, due to their more efficient implementations.

\vspace{1mm}
\noindent \textbf{Datasets.}
The datasets used in the experiments are listed in Table 2. We used 16 large real-world graphs with up to a billion edges. Networks are treated as undirected simple graphs, and are processed appropriately.

\vspace{1mm}\noindent \textbf{Settings.}
The tests are run on a 64 bit Linux machine with a Intel(R) Xeon(R) CPU E5-2650 v3 @ 2.30GHz processor, the L1, L2 and L3 cache of 32K, 256K, and 25600K respectively, with 591 GB of available RAM.

\begin{figure*}[thb]
	\center{
    \includegraphics[width=1\columnwidth]
	{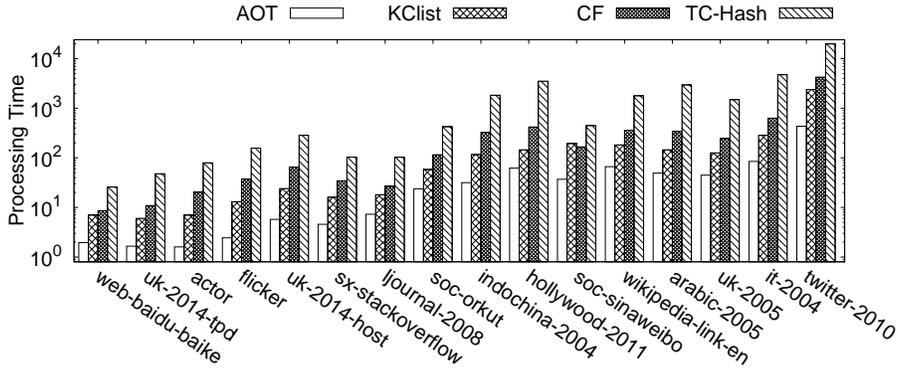}}
   \vspace{-3mm}
	\caption{\label{fig1} Performance Analysis}
    \label{exp:time_1}
\end{figure*}

\vspace{-2mm}
\subsection{Results against the State-of-the-art}
\vspace{-1mm}

Figure~\ref{exp:time_1} reports the relative running times of the algorithms tested. The measured time captures the elapsed time between the point when the graph is loaded until the point of successful program termination.
For the state-of-the-art methods, the kClist algorithm requires fewer running time than the CF algorithm.
For datasets containing up to 100 million edges, kClist is observed to significantly outperform CF; this gap in running time is less significant for graphs where the edge count is greater than 100 million.
There are also instances where CF is more efficient than kClist, as observed in the social-network \textit{soc-sinaweibo}.

In comparison, our algorithm AOT consistently outperforms the two state-of-the-art, which supports the tightened running bound of $\sum_{\{u,w\} \in E } min (deg^{+}(u), deg^{+}(v)) $ from its theoretical analysis.
We notice that on a large graph \textit{twitter-2010} that has $41.65$ million vertices, $1.2$ billion edges and contains $35$ billion triangles, the observed running times of kClist and CF are $2,381$ seconds and $4,230$ seconds respectively. In contrast, our algorithm listed all triangles in \textit{twitter-2010} in $433$ seconds and achieved a speedup of $10$-times.
It is noticed that hash-based CF (CF-Hash) is consistently outperformed by AOT 
with big margins, though two algorithms have the same asymptotic behavior.
This is because the high efficiency of look-up operation of the bitmap hash table
as well as the local ordering technique in AOT algorithm.
Recall that, without the adaptive orientation technique proposed in this paper,
hash-based CF cannot take this advantage. This reflects the non-trivial nature of our adaptive orientation technique.

\begin{figure*}[htb]
	\center{\includegraphics[width=1\columnwidth]
	{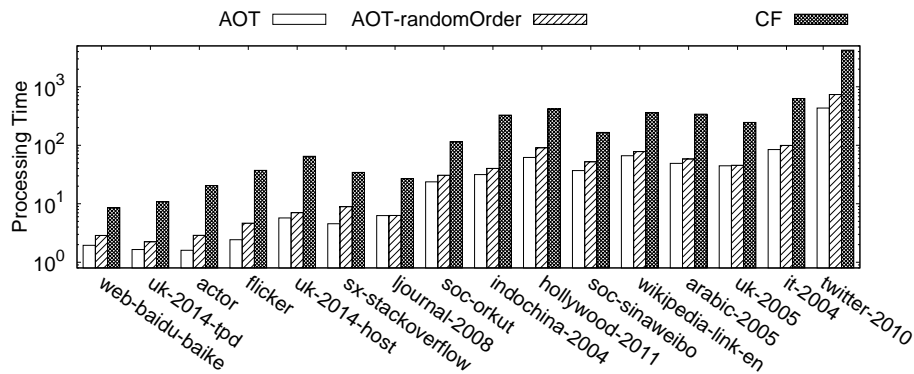}}
    \vspace{-3mm}
	\caption{\label{exp:fig2} Incremental Improvements}
\end{figure*}

\vspace{-2mm}
\subsection{More on AOT}
\vspace{-1mm}

To show the efficiency of our algorithm, we evaluate the benefits for having adaptive orientation and local ordering in our technique. For this setting, a baseline method is one that has neither adaptive orientation or local ordering.
We consider \cf algorithm as a proper baseline since it uses the existing orientation technique, but uses neither of the aforementioned techniques. In addition to considering the \aot algorithm with both adaptive orientation and local ordering.
We also require an algorithm that uses adaptive orientation without utilizing a local ordering technique,
for this, we consider our \aot algorithm with a random local ordering, denoted later as $AOT$-$randomOrder$.

As we can see in Figure~\ref{exp:fig2}, the processing time decreases after introducing adaptive orientation and the local ordering strategy.
In comparison to the baseline processing time, the adaptive orientation contributes a greater drop in processing time compared to that from the later adoption of the local-order strategy
Where the difference between $AOT$-$randomOrder$ and \cf is greater than that between $AOT$ and $AOT$-$randomOrder$.
This highlights that our adaptive orientation technique performs better than the orientation technique in its current state.
Furthermore, the results also show that using a local order reduces the running time needed on most graphs; this can be explained by an improvement in the algorithms cache performance.

\vspace{-2mm}
\subsection{Parallel Experiments}
\vspace{-1mm}

Our algorithm $AOT$ can be easily made parallel. This is achieved by processing vertices in parallel. As a result, we analyze the parallelism of our algorithm and compare it against the state-of-the-art methods.
$TC$-$Merge$ (i.e., parallel implementation of $CF$ proposed in~\cite{08tcstriangle}), $TC$-$Hash$ and $kClist$ all provided parallel implementations of their algorithms. For this parallel experiment, we consider the two largest datasets 
$It$-$2004$ and $Twitter$-$2010$.

\begin{figure}[htb]
\begin{center}
    \subfigure[It-2004]{
    \includegraphics[width=0.47\columnwidth]{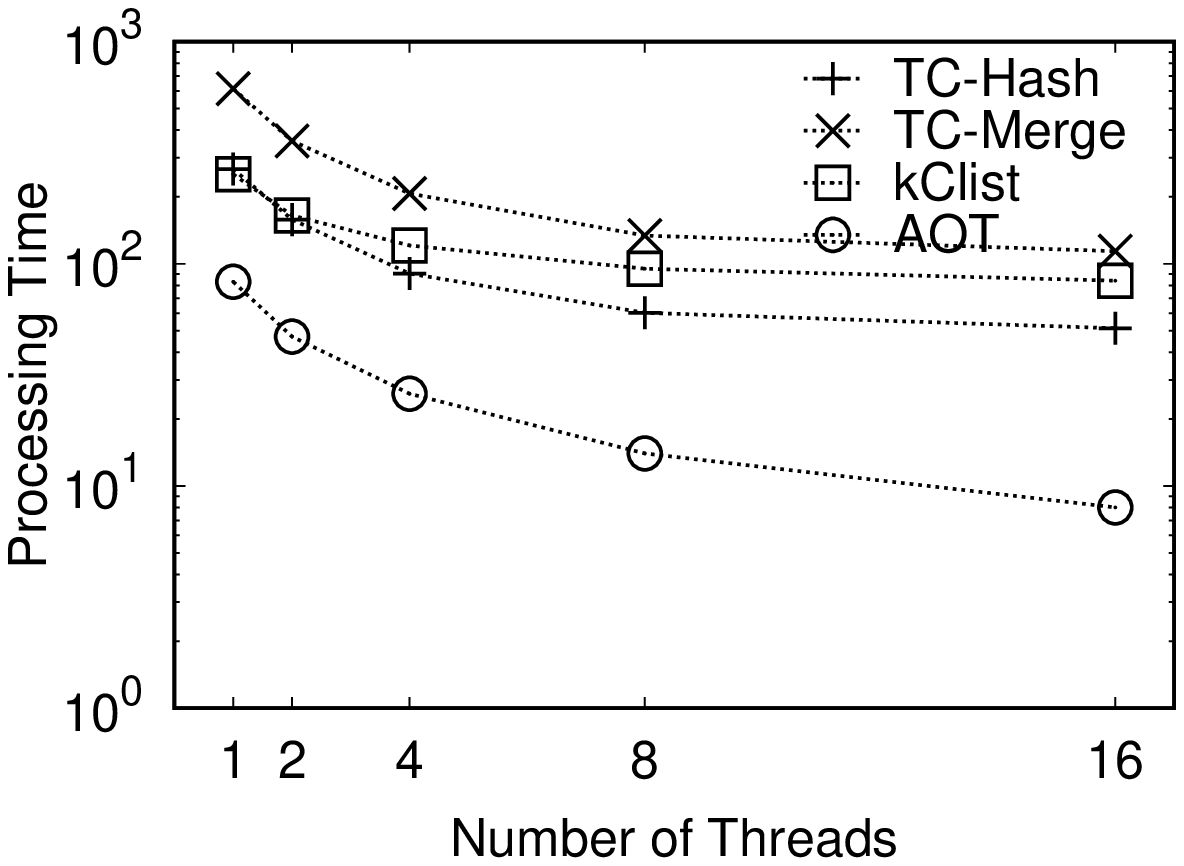}}
    \subfigure[Twitter-2010]{
    \includegraphics[width=0.47\columnwidth]{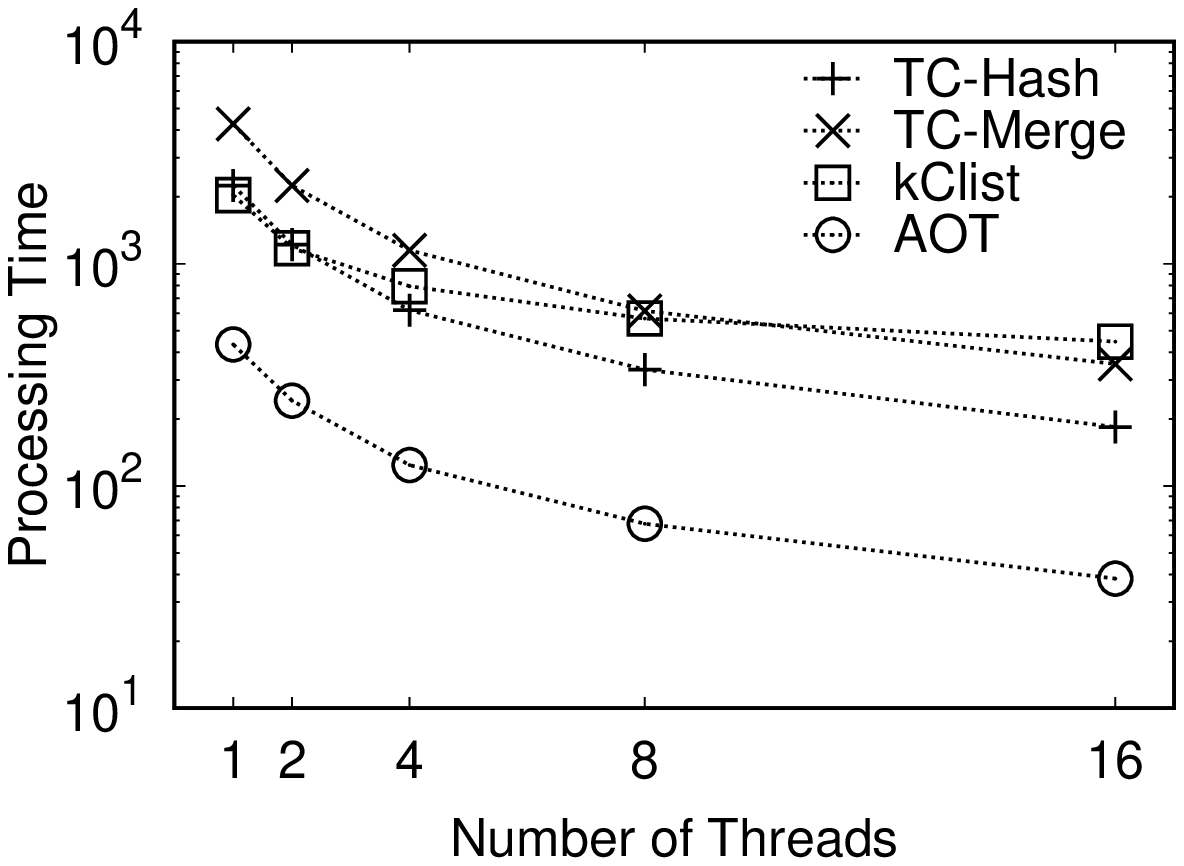}}
\end{center}
\caption{\small{Evaluating Parallel Performance}}
\label{fig:exp:parallel}
\end{figure}

As seen in Figure~\ref{fig:exp:parallel}, increasing the number of threads decreases the amount of processing time needed to list all triangles, this is expected and true for all four algorithms tested. 
In the case of $kClist$, the drop is less pronounced for both networks after $4$ threads.
In the case of $TC$-$Merge$ and $TC$-$Hash$, the drop in processing time is not visible when handling $It$-$2004$ past $8$ threads.
In contrast, this decrease is visible for our method $AOT$ across both datasets.
All in all, $AOT$ is consistently the fastest method in this parallel experiments.

\vspace{-4mm}
\section{Related Work}
\label{sec:rel}
\vspace{-2mm}

\vspace{1mm}\noindent \textbf{Triangle Listing}
In-memory algorithms for listing triangles have been extensively studied in the literature.
The edge-iterator~\cite{DBLP:journals/socnet/BatageljM01} and node-iterator~\cite{itai1978finding} are two popular triangle listing computing paradigms, which share the same asymptotic behavior~\cite{schank2005finding}.
A lot of subsequent algorithms are mostly improvements based on the original two paradigms.
While Ortmann was the first to formalize a generalized framework based on undirected graph orientation,
past literature Forward and Compact Forward(CF) had previously considered triangle-listing on induced directed graphs with respect to a vertex ordering ~\cite{schank2005finding}. In literature, the orientation technique is observed beyond triangle-listing; it is also applied for higher-order structure enumeration ~\cite{danisch2018listing}.
In more recent years, the topics of interest have shifted to parallel/distributed processing
(e.g.,\cite{shun2015multicore,park2016pte}),
efficient I/O external memory methods (e.g.,\cite{11KDDtriangle,14todstrianglelist}, and
the asymptotic cost analysis of triangle listing in random graphs~\cite{xiao2017asymptotic}.

\vspace{1mm}\noindent \textbf{Triangle Counting}
The triangle counting is a related problem to the triangle listing problem, solving the listing problem solves the counting problem. The triangle counting problem is the task to find the total number of triangles in a graph G. Compared to listing algorithms, counting algorithms find ways to compute the number without relying on the exploration of triangle instances.
Many algorithms have been designed to count triangles (e.g., \cite{08kddtriangle,DBLP:journals/pvldb/PavanTTW13,14trainglemapreduce}). 
Approximate methods are useful for settings that handle large-scale graphs, or settings where a given approximation is as useful as knowing the exact triangle count (e.g.,\cite{09KDDtriangle,turkoglu2017edge}). 

\vspace{-4mm}
\section{Conclusion}
\label{sec:con}
\vspace{-2mm}

The triangle listing is a fundamental problem in graph analysis with a wide range of applications.
This problem has been extensively studied in the literature. Although many efficient main memory algorithms based on the efficient orientation technique have been proposed,
in this paper, we pushed the efficiency boundary of the triangle listing and 
developed a new algorithm with best theoretical time complexity and practical performance.
On the theoretical side, we showed that our proposed algorithm has the time complexity
$\Theta(\sum_{ \langle u,v \rangle \in \vec{E} } \min\{ deg^{+}(u),deg^{+}(v)\}))$
where $\vec{E}$ is the directed edges in the oriented graph, which is the best known theoretical time complexity
for the problem of in-memory triangle listing so far.
On the practical side, our comprehensive experiments over $16$ real-life large graphs
show the superior performance of our \aot algorithm compared with two state-of-the-art techniques,
especially on large-scale graphs with billions of edges. 
\subsection*{Acknowledgement}
Lu Qin is is supported by ARC DP160101513. Ying Zhang is supported by FT170100128 and ARC DP180103096, Wenjie Zhang is supported by ARC DP180103096 and ARC DP200101116. Xuemin Lin is supported by 2018YFB1003504, ARC DP200101338, NSFC61232006, ARC DP180103096 and DP170101628.
\small{
\bibliographystyle{splncs04}
\bibliography{ref}}
\end{document}